\documentstyle [prb,aps,epsf]{revtex}
\begin{document}
\twocolumn[{
\widetext
\draft

\title{Violation of Kohler's rule
by the magnetoresistance\\
of a quasi-two-dimensional organic metal
}

\author{Ross H. McKenzie\cite{email}}

\address{School of Physics, University of New
South Wales, Sydney 2052, Australia}

\author{J. S. Qualls, S. Y. Han, and J. S. Brooks}

\address{
National High Magnetic Field Laboratory and
Department of Physics,
Florida State University, Tallahassee, FL 32306}

\date{Received October 17, 1997}

\maketitle
\mediumtext
\begin{abstract}
The interlayer magnetoresistance of the
quasi-two-dimensional metal
$\alpha$-(BEDT-TTF)$_2$KHg(SCN)$_4$
is considered.
In the temperature range from 0.5 to 10 K
and for fields up to 10 tesla
the magnetoresistance has a stronger
temperature dependence than the zero-field resistance.
Consequently Kohler's rule is not obeyed
for any range of temperatures or fields.
This means that the magnetoresistance cannot be described
in terms of semiclassical transport on a single Fermi surface
with a single scattering time.
Possible explanations for the violations
of Kohler's rule are considered,
both within the framework of semi-classical transport
theory and involving incoherent interlayer transport.
The issues considered are similar to those raised
by the magnetotransport of the cuprate superconductors.
\\
\\
To appear in Physical Review B, May 15, 1998.
\end{abstract}

\pacs{PACS numbers: 72.15.Gd, 74.70.Kn, 75.30.Fv, 71.45.Lr}

}] \narrowtext
Currently a great deal of attention
is being paid to the large magnetoresistance
of layered materials such as
magnetic multilayers\cite{multi} and 
manganese perovskites.\cite{kimura1}
This is motivated by potential applications
in magnetic recording and by the challenge of understanding
the physical origin of
the magnetoresistance, which  is very different
from that in conventional metals.\cite{pip} 
The magnetotransport of the metallic phase of
the cuprate superconductors also differs significantly
from conventional metals.\cite{ong,kimura2,and}
In this Rapid Communication we show that the magnetoresistance
of a particular organic metal may also be unconventional.

Layered organic molecular crystals
based on the\hfil\break 
bis-(ethylenedithia-tetrathiafulvalene) (BEDT-TTF)
molecule are model low-dimensional electronic systems.\cite{ish,wos}
 The family 
$\alpha$-(BEDT-TTF)$_2${\sl M}Hg(SCN)$_4$[{\sl M}=K,Rb,Tl]
 have a rich phase diagram depending on temperature,
 pressure, uniaxial stress, and magnetic field:
 metallic, superconducting, and density-wave phases are
possible.\cite{toyota,chen,campos}
Band structure calculations predict        co-existing
quasi-one-dimensional (open) and quasi-two-dimensional
(closed) Fermi surfaces.\cite{mori}
At ambient pressure these materials undergo a transition 
at a temperature $T_{DW} $  (8 K in the {\sl M} = K salt)
 into a low-temperature metallic  phase that has been argued
to be a density-wave (DW).
This phase is destroyed in high magnetic fields. 
There is
currently controversy as to whether this phase is a spin-density wave,
 a charge-density wave, 
or a mixture of both.\cite{toyota,pratt,ath,mck2,sasak}

The following picture of the low-temperature phase has been 
proposed.\cite{kar,iye} The nesting of the quasi-one-dimensional
Fermi surface
leads to a density-wave instability at $T_{DW} $.
Below $T_{DW} $ a gap opens on 
the quasi-one-dimensional
Fermi surface and the associated carriers no longer
contribute to the transport properties.
The density wave introduces a new periodic potential
into the system resulting in reconstruction
of the quasi-two-dimensional Fermi surface.
One of the proposed Fermi surface reconstructions
involves large open sheets.\cite{kar}
 Semi-classical transport theory can then explain
the large magnetoresistance and its angular dependence in the 
low-temperature phase.\cite{iye} The complete
field dependence of the resistance can also be explained
if magnetic breakdown is taken into account.\cite{mck}
However, in this paper we show
that the temperature dependence of the magnetoresistance
is inconsistent with the above picture.
In particular, the magnetoresistance is
shown to violate Kohler's rule,\cite{clo4}
raising issues similar to those considered
for the cuprate superconductors.\cite{ong,kimura2,and}

The temperature and field dependence  
of the magnetoresistance 
of many metals can be analysed in terms
of Kohler's rule.\cite{pip}
Semiclassical transport theory 
based on the Boltzmann equation predicts Kohler's rule to hold if
there is a single species of charge carrier and the 
scattering time $\tau$ is the same
at all points on the Fermi surface.
The dependence of the resistance on the field is then
contained in the quantity $\omega_c \tau$
where $\omega_c$ is the frequency at which
the magnetic field $B$ causes
the charge carriers to sweep across the Fermi surface.
Since the resistance in zero field  is proportional to
the scattering rate, the field dependence
of the magnetoresistance of samples with
different scattering times (either due
to different purity or temperature $T$)
can be related by rescaling the field
by the zero-field resistance $R(0,T)$:
\begin{equation}
{R(B,T) \over R(0,T)} = F (\omega_c \tau) =
f \left({B \over R(0,T)} \right) \label{kohl}
\end{equation}
This is Kohler's rule and the corresponding
plots are known as Kohler plots.
It holds regardless of the topology and geometry
of the Fermi surface.

Resistance measurements were performed on a 
 single crystal of
$\alpha$-(BEDT-TTF)$_2$KHg(SCN)$_4$ 
using a standard four-wire AC technique with a 10 microamp
current along the {\bf b} axis
(the least conducting axis).
 Sample contacts were made on the faces of the {\bf a-c}
planes
with 12.5 micrometer gold wire attached via carbon paint. The magnetic
field was applied parallel to the {\bf b} axis.
 Measurements were performed in a
$^3$He cryostat using a 33 tesla Bitter magnet at the National High Magnetic
Field Laboratory in Tallahassee.
Fig.   1 shows the field dependence of the 
interlayer resistance 
 at several different temperatures.
The magnetic field is parallel to the
current and perpendicular to the layers.
The data is consistent with previously
published data on this class of
materials.\cite{sasak,mck,brooks,sasaki,sasaki2}

Given that the current direction and
magnetic field are parallel one expects
no Lorentz force on the electrons.
This raises the question of
the origin of such a large longitudinal 
magnetoresistance.
Semi-classical theories explain this by
assuming that the interlayer hopping
also involves a substantial simultaneous
hopping parallel to the layers.\cite{iye}
Hill has shown how such hopping, and the
associated warping of the Fermi surface
in the interlayer direction, can be used
to explain cyclotron resonance experiments.\cite{hill}
The microscopic justification
for assuming this type of interlayer hopping is not clear.

The strong angular dependence of
the interlayer magnetoresistance\cite{chen,kar,sasaki,sing}
 implies that it is predominantly orbital in
origin. When the field is parallel to the layers
or at certain magic angles the magnetoresistance is
several times smaller than when
the field is perpendicular to the layers. 
If the magnetoresistance was predominantly due
to the field coupling to the spins it should be
almost isotropic.

Fig. 2 shows a Kohler plot of the data in Fig.  
1 as well as data at additional temperatures.
It covers fields up to about 10 tesla.
If Kohler's rule held all of the curves would
collapse onto a single curve.
They do not because the magnetoresistance
varies strongly with temperature
but the zero-field resistance is only
weakly temperature dependent (Fig. 1).
Note that there is {\it no} field range over 
which Kohler's rule holds. This rules out
explaining the deviation in terms of
quantum effects or magnetic breakdown.

We now consider five 
possible explanations for the violation
of Kohler's rule, within the framework of
semi-classical transport theory.
(i) The electronic structure varies with
temperature due to formation of the density wave.
This can explain the temperature dependence
between 4 K and 10 K.
However, in density wave systems the electronic
energy gap varies very little at temperatures
less than half the transition temperature.\cite{gru}
In this system, below 4 K, there is little change
in the zero-field  resistance (see Fig. 1),
Hall resistance,\cite{sasaki2}    Knight shift,
and nuclear magnetic relaxation rate, $ 1/(T_1 T)$.\cite{kanoda}
This suggests that the electronic structure and density of
states does not vary significantly below 4 K
and so cannot explain the large temperature
dependence of the magnetoresistance.\cite{house2}

(ii) There is more than one type of carrier and
their mobilities have different temperature dependences.
The existence of more than one type of carriers
in the low-temperature phase is 
suggested by the observation of 
more than one magneto-oscillation frequency\cite{uji1}
and more than one cyclotron resonance frequency.\cite{ohta}
To illustrate how this can lead to
violations of Kohler's rule we
consider the case of two carriers.\cite{newson}
Let $n_1$ and $n_2$ denote the densities
and $\mu_1$ and $\mu_2$ denote the mobilities
of the carriers.
The zero-field resistance is $\rho_0 = (n_1 \mu_1 + n_2 \mu_2)^{-1}$.
At low fields the transverse\cite{caveat} magnetoresistance
is\cite{pip}
\begin{equation}
{\Delta\rho_{xx} \over \rho_0} =
 {n_1 n_2 \mu_1 \mu_2 
 (\mu_1 - \mu_2)^2 B^2
\over (n_1 \mu_1 + n_2 \mu_2)^2}
\label{two}
\end{equation}
Hence, if $\mu_1$ and $\mu_2$ have a
different temperature dependence so
will the resistance and magnetoresistance.
To see this clearly    
consider the particular case where $n_1 \sim n_2$ and $ \mu_1 \gg \mu_2$
then $ \rho_0 \simeq (n_1 \mu_1)^{-1}$
and $ {\Delta\rho_{xx} \over \rho_0} \simeq 
 { n_2 \mu_1 \mu_2 \over n_1 } B^2.$
Hence, if $\mu_2$ has a much stronger temperature dependence
than $\mu_1$ then the desired behavior is obtained.\cite{caveat0}
However, in this limit the Hall resistance is
$R_H \simeq \mu_2/(n_1 \mu_1)$
and so should be strongly temperature dependent.
However, this is inconsistent with observations
(albeit on a different sample).\cite{sasaki2,top}

(iii) The temperature dependence of the
scattering  rate varies significantly at
different points on the Fermi surface.\cite{conv}
Similar ideas about ``hot spots'' have 
been proposed to explain the magnetotransport
in the cuprates\cite{pines} and quasi-one-dimensional
organic metals\cite{chaikin,yak}.
A different temperature dependence for the
resistance and magnetoresistance arises
because the former is related to the inverse of the
average of the scattering time over the Fermi surface
and the latter (at high fields) is related to the
average of the scattering rate over the Fermi surface.\cite{yak}
Alternatively, the magnetoresistance can be shown
to be the variance of the Hall angle over the
Fermi surface.\cite{ong}
The non-uniform scattering rate also leads to
a temperature dependence of the Hall resistance $R_H$ since
it is given by\cite{hall0}
\begin{equation}
R_H = { 1 \over n e} { \langle \tau^2 \rangle \over \langle \tau
\rangle^2} 
\label{hall}
\end{equation}
where $\langle ... \rangle$ denotes an average over the Fermi surface.
However, again this explanation requires
the Hall resistance  to vary significantly below 4 K,
whereas it does not.\cite{sasaki2,top}

(iv) The scattering times associated with the
magnetoresistance and the zero-field resistance
are distinct and have different temperature
dependences. This hypothesis\cite{and} has
been proposed to explain the
unusual temperature dependence of the magnetotransport
(including the violation of Kohler's rule)
 in the metallic phase of the cuprate superconductors.\cite{ong,kimura2}
Distinct scattering times are associated with the decay
of electric and Hall currents and denoted $\tau_0$ and $\tau_H$,
respectively. The zero-field conductivity $\sigma_{xx}(0) \sim \tau_0$,
the magnetoconductivity $\sigma_{xx}(B)- \sigma_{xx}(0) \sim 
B^2 \tau_0 \tau_H^2$,\cite{caveat}
and the Hall conductivity $\sigma_{xy} \sim B \tau_0 \tau_H$.
Consequently, this explanation also requires the
Hall resistance of
$\alpha$-(BEDT-TTF)$_2$KHg(SCN)$_4$ 
to vary significantly below 4 K.
Measurements 
suggest that it does not.\cite{sasaki2,top}

(v) The scattering time $\tau$ is field dependent
in a way that $\tau(B,T)/\tau(0,T)$ is
temperature dependent.
Several calculations have considered the electron-electron
scattering rate in the quasi-one-dimensional Bechgaard 
salts (TMTSF)$_2$X and suggested that it is field dependent.\cite{lebed}
Alternatively, if the scattering is due to local magnetic moments,
possibly associated with a spin-density wave, then
that will vary with field.\cite{kimura1}
Although these explanations for the violation
of Kohler's rule are possible it should be stressed that
if they are correct then the origin of the magnetoresistance
in these materials is quite
different from what has been proposed.\cite{kar,iye}


The possible failure of semi-classical transport theory
to describe the interlayer magnetoresistance
raises the question:
is the interlayer transport incoherent, i.e.,
does the concept of Bloch states (on which
the Boltzmann equation depends) have meaning?

For this class of materials
Yoshioka\cite{yosh} has proposed an
explanation for the magnetoresistance and
its angular dependence that does not 
involve coherent interlayer transport.
Yoshioka's model
assumes that there is a periodic potential due
to a density wave in each layer.
A magnetic field then produces a periodic  potential
whose period
along the $\bf b$ axis, i.e, perpendicular to the layers,
is {\it incommensurate} with the interlayer spacing.
If the magnitude of this potential is larger than
the interlayer hopping rate than all
the states along the $\bf b$ axis will be localized.\cite{sokol}
The strength of the incommensurate  potential
increases with field and makes the states more localized.
Hence, the interlayer resistance increases with increasing field.
The incommensurability of the potential varies as the field
is tilted. At certain angles the potential will 
become commensurate, the states will no longer
be localised and the magnetoresistance will vanish.
The model correctly predicts these angles.\cite{yosh}
Although all the states are localized the conductivity
should be non-zero at finite temperature due to variable range hopping.
As the temperature is lowered variable range hopping 
becomes harder and resistance goes up.\cite{vrh}
Hence, in this model the temperature dependence
of the magnetoresistance is unrelated to that of
the zero-field resistance and 
Kohler's rule would not be expected to hold.
However, this model would predict that,
contrary to what is observed, the magnetoresistance
does not saturate as the temperature is lowered.\cite{weiss}

The issue of incoherent interlayer transport
has recently been considered for the
cuprate superconductors\cite{cuprate,str}
and for the layered organic crystal      (TMTSF)$_2$PF$_6$, 
which under pressure is a quasi-one-dimensional metal.
Its magnetoresistance strongly violates Kohler's
rule and only depends on the component
of magnetic field perpendicular to the layers.\cite{danner}
Although there are some similarities there are also differences
to the material studied here. For example,
in (TMTSF)$_2$PF$_6$, the angular dependence
of the magnetoresistance
has a minimum when the field is perpendicular to
the layers whereas for 
$\alpha$-(BEDT-TTF)$_2$KHg(SCN)$_4$ 
it is a maximum.\cite{chen}
Although it would be interesting to apply the ideas
in Ref. \onlinecite{str} to the
data presented here it is not clear how to do so.


In conclusion,
the temperature dependence of the interlayer
magnetoresistance of the
quasi-two-dimensional metal
$\alpha$-(BEDT-TTF)$_2$KHg(SCN)$_4$
cannot be explained in terms of
existing theoretical models including,
(i) semi-classical transport on a single Fermi surface
with a single scattering time\cite{iye} and (ii)
Yoshioka's model\cite{yosh} involving
incoherent interlayer transport .
We suggest several directions for future work.
Experimentally,  Kohler's rule should be
tested outside the low-temperature phase
and in other metals
based on the BEDT-TTF molecule.
Hall resistance  and magnetoresistance
measurements should be done on the
same  sample to completely rule out the
``hot spot'' and
two scattering time hypotheses for these systems.
Theoretically, we need calculations of the magnetoresistance for
models\cite{cuprate,str} involving incoherent interlayer transport.


We thank S. Hill, B. Kane, A. G. Lebed and J. Singleton
for helpful discussions.
Work at UNSW was supported by the Australian Research Council.
JSQ, SYH, and JSB were supported in part by NSF grant DMR 95-10427.
Work at the National High Magnetic Field Laboratory
was supported by NSF Cooperative Agreement
No. DMR-9016241 and the state of Florida.

\newpage
\begin{figure}
\centerline{\epsfxsize=9.3cm 
 \epsfbox{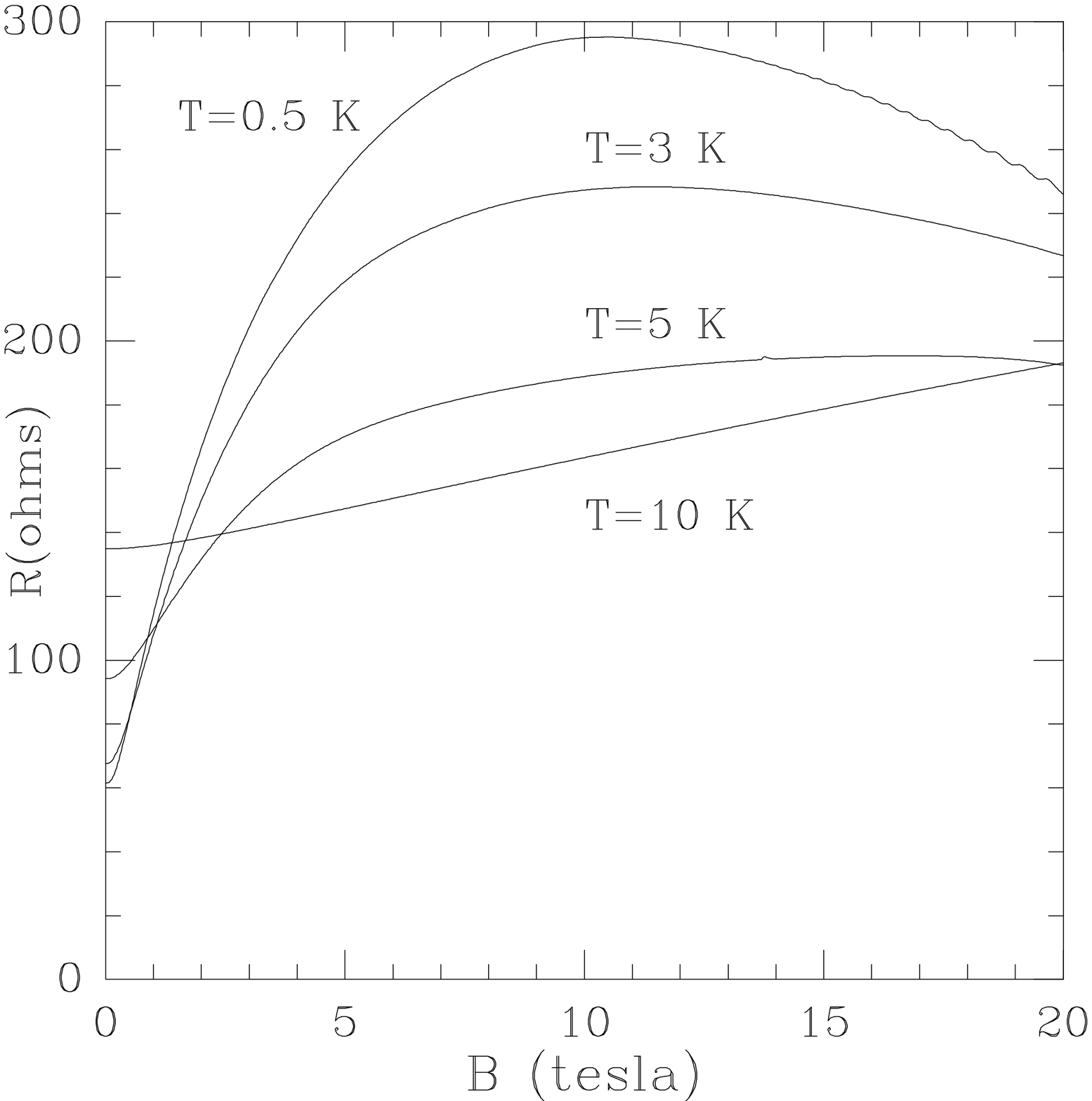}}
\vskip 0.5cm
\caption{
Magnetic field dependence of the interlayer resistance of 
$\alpha$-(BEDT-TTF)$_2$KHg(SCN)$_4$ 
 at several temperatures.
The magnetic field and the current direction were
perpendicular to the layers, i.e., parallel to the
least-conducting direction.
\label{fig1}}
\end{figure}

\begin{figure}
\centerline{ \epsfxsize=9.3cm \epsfbox{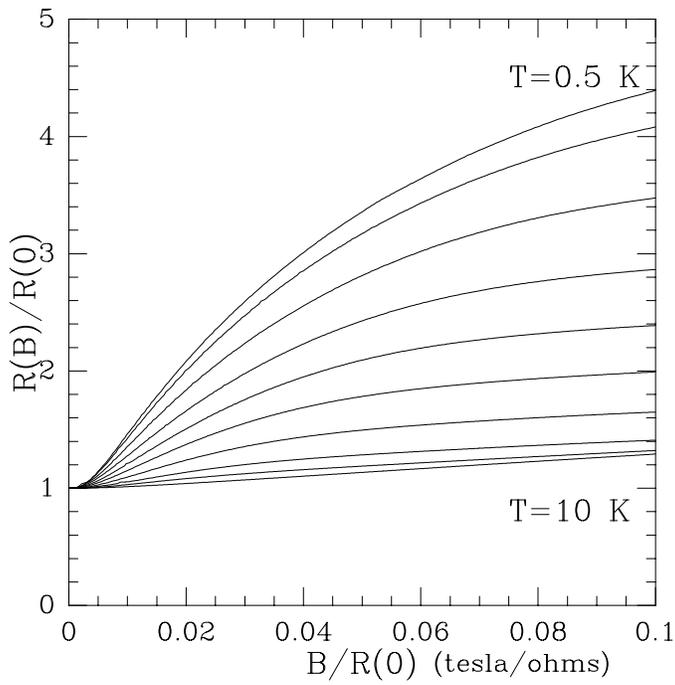}} 
\vskip 0.5cm
\caption{
Kohler plot of the magnetoresistance.
The temperatures of the curves shown are (from top
to bottom) 0.5, 1.5, 3.0, 3.5, 4.2, 5.0, 6.0,
7.0, 8.0, and 10.0 K.
If Kohler's rule held then all the curves would lie
on top of on one another.
\label{fig2}}
\end{figure}


\begin{references}
\bibitem{house2} On the other hand,
in Ref. \protect\onlinecite{sing} it
was found that the relative intensity of 
the magneto-oscillations of different frequency
varied between 0.7 K and 4 K. 
Although this suggests minor modification of
the Fermi surface  it is not clear
how this could be responsible for the violation
of Kohler's rule.
\bibitem{caveat} This expression is for the 
transverse magnetoresistance whereas our measurements
were for the longitudinal magnetoresistance.
In order for this expression to be relevant
to the experiments we must assume the interlayer hopping involves a 
component parallel to the layers. Then one can
anticipate the longitudinal conductivity
being described by an expression similar to this.
\bibitem{caveat0}
Note that in order for the magnetoresistance to
be of order one at a field of one  tesla,
as is observed, this explanation requires that
the mobilities must be sufficiently large that
the scattering times are less than $10^{-11}$ seconds.
This is comparable to values for the highest mobility
semiconductor heterostructures.
\bibitem{conv}
The variation of the scattering rate
over the Fermi surface leads to the violation of Kohler's rule
in conventional metals at high temperatures where
the scattering is dominated by small angle scattering
by phonons. (See p. 126 of Ref. \onlinecite{pip}).
Yet Kohler's rule is still obeyed 
at low fields, and the magnetoresistance
increases with increasing temperature.
That is quite different from the deviations observed here.
\end{references}
\end{document}